\documentclass[epj,nopacs]{svjour} 
\usepackage[]{amsmath}
\usepackage[]{amssymb}
\usepackage[]{amsfonts}
\usepackage{graphics}
\begin{document}
\title{\LARGE{Analysing powers for the reaction 
\boldmath $\vec{\rm n} {\rm p} \rightarrow {\rm p} {\rm p} \pi^{-}$ \\
and for np elastic scattering from 270 to 570 MeV}}
\titlerunning{Analysing powers for the reaction 
$\vec{\rm n} {\rm p} \rightarrow {\rm p} {\rm p} \pi^{-}$ \ldots}

\author{
M.~Daum\inst{2} \and
M.~Finger\inst{3}\fnmsep\inst{4} \and
M.~Finger, Jr.\inst{4} \and
J.~Franz\inst{1} \and
F.H.~Heinsius\inst{1} \and
A.~Janata\inst{4} \and
K.~K\"onigsmann\inst{1} \and
H.~Lacker$^{\ast}$\inst{1} \and
F.~Lehar\inst{5} \and
H.~Schmitt\inst{1} \and
W.~Schweiger\inst{1} \and
P.~Sereni\inst{1} \and
M.~Slune\v cka\inst{3}\fnmsep\inst{4}
}


\institute{Fakult\"at f\"ur Physik der Universit\"at Freiburg, D-79104 
Freiburg, Fed. Rep. Germany \and 
PSI, Paul-Scherrer-Institut, CH-5232 Villigen, 
Switzerland \and 
Charles University, Faculty of Mathematics and Physics, 
V Hole\v sovi\v ck\'ach 2, CZ-18000 Praha 8, Czech Republic \and 
Joint Institute for Nuclear Research, LNP, RU-141980 Dubna, Moscow Region, Russia \and 
DAPNIA/SPP, CEA/Saclay, F-91191 Gif-sur-Yvette CEDEX, France}
\mail{lacker@lal.in2p3.fr}

\abstract{
The analysing power of the reaction
${\rm n}{\rm p} \rightarrow {\rm p}{\rm p} \pi^{-}$ for neutron 
energies between threshold and 570 MeV has been determined using 
a transversely polarised neutron beam at PSI. The reaction has 
been studied in a kinematically complete measurement using a 
time-of-flight spectrometer with large acceptance. Analysing powers 
have been determined as a function of the c.m. pion angle in different 
regions of the proton-proton invariant mass. They are compared to 
other data from the reactions 
${\rm n}{\rm p} \rightarrow {\rm p}{\rm p} \pi^{-}$ 
and
${{\rm p}{\rm p} \rightarrow {\rm p}{\rm p} \pi^{0}}$.
The np elastic scattering analysing power was determined as a 
by-product of the measurements.
}
\date{Received: YYYY ?, 2001 / Accepted: XXXX ?, 2001}

\maketitle

\section{Introduction}

\noindent{
Single pion production is the main inelastic hadronic process 
in nucleon-nucleon collisions for beam energies below 1 GeV. 
During the last decade, new results have been obtained for 
proton-proton induced reactions. The new data, taken at proton 
cooler synchrotrons, triggered new theoretical efforts. However, 
a complete understanding of the various production mechanisms 
requires high quality data in all possible pion production 
reactions. In this paper, we report on measurements of the 
spin dependence in the reaction 
$\vec{\rm n}{\rm p} \rightarrow {\rm p}{\rm p} \pi^{-}$ 
and in the elastic np scattering.
}

\subsection{Pion production in neutron proton collisions}

\noindent{
Assuming isospin invariance in strong interactions, all single 
pion production reactions in nucleon-nucleon collisions with 
three body final states can be decomposed into three partial 
cross sections $\sigma_{I_{i}I_{f}}$\,\cite{ROS1}. Here, $I_{i}$ 
and $I_{f}$ denote the isospin of the nucleon-nucleon system in 
the initial and final state. Information on the isoscalar cross 
section $\sigma_{01}$ can be obtained by comparing charged pion 
production in neutron-proton collisions with the reaction 
${\rm p}{\rm p} \rightarrow {\rm p}{\rm p} \pi^{0}$\,\cite{DAU1}.
}

\noindent{
The production mechanism is often discussed in terms of partial 
waves. In this paper, the notation 
\begin{displaymath}
^{2S+1}\!L_{J} \rightarrow ^{2S'+1}\!\!\!L'_{J'}\ell_{J} 
\end{displaymath}
is adopted, where $S$ is the total spin, $L$ the orbital angular 
momentum and $J$ the total angular momentum of the two nucleons in 
the intial state, while $S'$, $L'$ and $J'$ give the corresponding 
angular momenta in the final state. The orbital angular momentum of 
the pion with respect to the final state nucleon-nucleon system is 
denoted $\ell$.
}

\noindent{
Partial waves with a relative angular momentum $L=0$ for the 
proton-proton final state play a particular role due to the 
strong final state interaction at small relative momenta. 
Amongst the partial waves with pp($^{1}{\rm S}_{0}$) final 
states,
$^{3}{\rm P}_{0} \rightarrow ^{1}\!{\rm S}_{0}{\rm s}_{0}$, 
$^{3}{\rm P}_{2} \rightarrow ^{1}\!{\rm S}_{0}{\rm d}_{2}$
and
$^{3}{\rm F}_{2} \rightarrow ^{1}\!\!{\rm S}_{0}{\rm d}_{2}$
contribute to $\sigma_{11}$, whereas 
$^{3}{\rm D}_{1} \rightarrow ^{1}\!\!{\rm S}_{0}{\rm p}_{1}$ 
and
$^{3}{\rm S}_{1} \rightarrow \,^{1}{\rm S}_{0}{\rm p}_{1}$ 
contribute to $\sigma_{01}$. Effects from these partial waves 
are enhanced if the phase space is restricted to small 
proton-proton invariant masses.
}

\subsection{Former Experiments}

\noindent{
For a long time, the experimental knowledge on the reaction 
${\rm n}{\rm p} \rightarrow {\rm N}{\rm N} \pi^{\pm}$ was rather 
weak. Many observables, e.g., invariant mass or angular distributions, 
as well as integrated cross sections were not very well 
known\,\cite{HAND1,RUSH1,DZH1,KLE1,THO1,DAK1,TSU1,KAZ1,BAN1,BAC1}.
As a consequence, no conclusive result concerning the size and the role 
of the cross section $\sigma_{01}$ has been found, see {\it e.g.} 
Refs.\,\cite{VER1,BYS1}. In ref.\,\cite{VER1} one strictly assumes isospin 
invariance, whereas in ref.\,\cite{BYS1} one compares the results of this 
assumption with the calculations, where np inelastic one-pion-production 
reactions are considered to be independent of each other. 
For a discussion of the partly contradictory experimental results for
the reactions ${\rm n}{\rm p} \rightarrow {\rm N}{\rm N} \pi^{\pm}$ 
and ${\rm p}{\rm p} \rightarrow {\rm p}{\rm p} \pi^{0}$ see also 
Ref.\,\cite{DAU1}.
}

\noindent{
During the last decade, new medium energy accelerators 
provided secondary neutron beams of high intensity and 
polarisation. This resulted in a substantial improvement 
of the experimental situation. Single spin observables 
for the reaction 
${\rm n}{\rm p} \rightarrow {\rm p}{\rm p} \pi^{-}$ have 
been measured at TRIUMF at 443 MeV\,\cite{BAC1} and at 
SATURNE at 572, 784, 1012 and 1134 MeV\,\cite{TER1}. 
Exclusive experiments were performed at TRIUMF, with 
proton beam energies of 353, 403 and 440 MeV incident 
on a deuterium target\,\cite{DUNC1,DUNC2}.
Events with small proton-proton invariant masses were
selected to investigate partial wave contributions with
a pp($^{1}{\rm S}_{0}$) final state. The results showed 
the significance of the $\sigma_{01}$ cross section in 
that particular phase space configuration. 
A partial wave analysis considering 
$^{3}{\rm S}_{1} \rightarrow \,^{1}{\rm S}_{0}{\rm p}_{1}$ 
and 
$^{3}{\rm D}_{1} \rightarrow \,^{1}{\rm S}_{0}{\rm p}_{1}$ 
for the $I=0$ and 
$^{3}{\rm P}_{0} \rightarrow \,^{1}{\rm S}_{0}{\rm s}_{0}$ 
for the $I=1$ initial state was performed\,\cite{DUNC2}. 
At 440 MeV, a small contribution from pion d-waves, 
$^{3}{\rm P}_{2} \rightarrow \,^{1}{\rm S}_{0}{\rm d}_{2}$ 
and 
$^{3}{\rm F}_{2} \rightarrow \,^{1}{\rm S}_{0}{\rm d}_{2}$, 
has been reported\,\cite{DUNC2}. 
}

\noindent{
Recently, differential and integrated cross sections for the 
reaction ${\rm n}{\rm p} \rightarrow {\rm p}{\rm p} \pi^{-}$ 
between threshold and 570 MeV have been measured at PSI\,\cite{DAU1}.
All observables revealed a significant contribution of $\sigma_{01}$. 
An enhancement in the proton-proton invariant mass distribution was 
observed at small values in agreement with the expected signal from Sp 
partial waves. In addition, large anisotropies and forward-backward 
asymmetries in the pion angular distributions were observed which was 
interpreted as a strong contribution of partial waves with Sp final 
states. From the measurement of the 
${\rm n}{\rm p} \rightarrow {\rm p}{\rm p} \pi^{-}$ cross section, 
$\sigma_{01}$ was obtained using existing 
${\rm p}{\rm p} \rightarrow {\rm p}{\rm p} \pi^{0}$ data
\,\cite{MEY1,BON1,RAP1,STA1,DUN1}. 
It was found to be of the same order as $\sigma_{11}$ in 
the energy range between 315 MeV and 400 MeV. 
The excitation function of $\sigma_{01}$ is reasonably 
described by a function $\propto \eta^{4}$, where 
$\eta=p_{\pi,{\rm max}}^{*}/m_{\pi^{+}}$ is the maximum 
value of the dimensionless c.m. pion momentum. This 
dependence is expected if $\sigma_{01}$ is carried by 
Sp partial waves.
}

\noindent{
Within the above mentioned PSI experiment\,\cite{DAU1} 
also spin dependent observables of the reaction 
$\vec{\rm n}{\rm p} \rightarrow {\rm p}{\rm p} \pi^{-}$ 
have been measured and are presented in this paper. 
}

\noindent{
Spin dependent observables are very sensitive to the 
interference between amplitudes from the $I=0$ and $I=1$
initial state. Hence, the measurement of spin observables 
for the reactions 
${\rm n}{\rm p} \rightarrow {\rm p}{\rm p} \pi^{-}$ and
${\rm p}{\rm p} \rightarrow {\rm p}{\rm p} \pi^{0}$ can
provide additional information on $\sigma_{01}$.
}

\section{Experiment}

\noindent{
The experiment was performed at the Paul-Scherrer-Insti-\\
tut (PSI).
The set-up and the analysis are described in more detail 
elsewhere\,\cite{DAU1,LAC1}. Data were taken with a transversely 
polarised neutron beam with about $50~\%$ of the data measured 
with horizontal and about $50~\%$ with vertical polarisation. 
To minimize detector induced asymmetries, the polarisation of 
the primary proton beam was reversed every second.
}

\noindent{
The proton beam polarisation was monitored by measuring the
rate asymmetry between both polarisation directions for protons 
elastically scattered on a thin Carbon target\,\cite{ARN1}.
The energy dependent neutron beam polarisation was measured 
in a former experiment\,\cite{ARN2}.
}

\noindent{
Two beam monitors\,\cite{ARN1} were used to record continously 
the polarised neutron beam properties during data taking. This
information was used in the off-line analysis to check possible
drifts of the beam polarisation and for eventual intensity and
position differences for the two polarisation directions.
}

\noindent{
For the kinematically complete measurement of the reaction 
${\rm n}{\rm p} \rightarrow {\rm p}{\rm p} \pi^{-}$, a 
time-of-flight (TOF) spectrometer with large angular and momentum 
acceptance was used. It consisted of a liquid hydrogen target, 
two drift chamber stacks together with a two dimensional 
scintillator hodoscope and a $3 \times 3~{\rm m}^2$ TOF wall. 
Events were selected by requiring at least two hits in the 
hodoscope, at least one hit in the TOF wall and hits in the 
first drift chamber. In addition, events fulfilling a minimum 
bias trigger were selected with a prescaling factor to study 
elastic scattering events. The experiment relied on the measurement 
of the energy for the incident neutron and the emission angles 
and velocities of at least two of the three charged particles 
in the final state. The energy of each incident neutron was
determined from a TOF measurement along a 20 m long flightpath 
using the 50 MHz time structure of the neutron beam.
The reaction ${\rm n}{\rm p} \rightarrow {\rm p}{\rm p} \pi^{-}$ 
was reconstructed using a kinematical fit technique. Background 
from the target surroundings in the final data sample was 
measured with an empty target cell and was found to be between 
8~\% at 315 MeV and 4~\% at 550 MeV. Monte Carlo simulation 
studies showed that background from other reactions in the 
liquid hydrogen target were negligible. For details of the event 
identification, see ref.\,\cite{DAU1}.
}

\section{Determination of analysing powers}

\noindent{
For a given neutron energy $T_{\rm n}$, the following basis in the c.m. 
system of the reaction ${\rm n}{\rm p} \rightarrow {\rm p}{\rm p} \pi^{-}$,
$\{\vec{S},\vec{N},\vec{L}\}$, is chosen: 
The unit vector $\vec{L}$ is defined by the neutron momentum in 
the c.m. system $\vec{L}=\vec{p_{\rm n}^{*}}/|\vec{p_{\rm n}^{*}}|$; 
$\vec{N}$ is the vector normal to the reaction plane defined by 
$\vec{N}=(\vec{p_{\rm n}^{*}}\times 
\vec{p_{\pi}^{*}})/|\vec{p_{\rm n}^{*}}\times \vec{p_{\pi}^{*}}|$;
$\vec{S}$ is chosen such that a right-handed orthonormal system 
is obtained. 
In the present experiment, the neutron beam is polarised, the target 
is not polarised and the polarisation of the final state protons is 
not analysed. In this case, the spin dependent cross section 
$d \sigma$ reads
\begin{eqnarray}
d \sigma 
= d \sigma_{0} \cdot 
(1 + P_{S}\cdot A_{S0} + P_{N}\cdot A_{N0} + P_{L}\cdot A_{L0})
\end{eqnarray}
where $d \sigma_{0}$ is the spin-averaged differential cross 
section and $P_{S}$, $P_{N}$ and $P_{L}$ are the projections of 
the beam polarisation vector $\vec{P}$ onto the three basis vectors 
$\vec{S}$, $\vec{N}$ and $\vec{L}$. The observables $A_{S0}$, $A_{N0}$ 
and $A_{L0}$ are called (beam) analysing powers. 
}

\noindent{
For a fixed neutron energy, the cross section $d \sigma$ is a 
function of five independent kinematical variables in the final 
state. Integrating over all phase space variables except the 
proton-proton invariant mass $M_{\rm pp}$, the pion c.m. angle 
$\theta_{\pi}^{*}$ and the angle $\phi$ between $\vec{N}$ and 
$\vec{P}$, one obtains for a transversely polarised neutron beam
\begin{eqnarray}
\begin{array}{c}
\!\!\!\!\!\!\!\!\!\!\!\!\!\!\!\!\!\!\!\!\!\!\!\!\!\!\!\!\!\!\!\!
\!\!\!\!\!\!\!\!\!\!\!\!\!\!\!\!\!\!\!\!\!\!\!\!\!\!\!\!\!\!\!\!
\!\!\!\!\!\!\!\!\!\!\!\!\!\!\!\!\!\!\!\!\!\!\!\!\!\!\!\!\!\!\!
d \sigma(T_{\rm n}, M_{\rm pp}, \theta_{\pi}^{*}, \phi) = \\
\!\!\!\!\!\!\!\! 
d \sigma_{0}(T_{\rm n}, M_{\rm pp}, \theta_{\pi}^{*})\cdot 
(1+P(T_{\rm n})\cdot A_{S0}(T_{\rm n}, M_{\rm pp}, \theta_{\pi}^{*})\cdot \sin{\phi} \\
\,\,\,\,\,\,\,\,\,\,\,\,\,\,\,\,\,\,\,\,\,\,\,\,\,\,\,\,\,\,\,\,
\,\,\,\,\,\,\,\,\,\,\,\,\,\,\,\,\,\,\,\,
+\,P(T_{\rm n})\cdot A_{N0}(T_{\rm n}, M_{\rm pp}, \theta_{\pi}^{*})\cdot \cos{\phi})~,
\end{array}
\end{eqnarray}
since the longitudinal polarisation component vanishes, $ P_{L}=0$. 
}

\noindent{
The analysing powers $A_{N0}$ and $A_{S0}$ were determined 
using the method of weighted sums\,\cite{BES1} which assumes 
an azimuthal symmetry of the detector around the beam axis. The 
assumption of parity conservation in strong interactions implies 
$A_{S0}(M_{\rm pp}, \theta_{\pi}^{*})=0$. Hence, the measurement 
of $A_{S0}$ provides an important cross-check for the analysis.
}

\noindent{
The beam polarisation depends on the neutron energy\,\cite{ARN2}. 
It was taken into account in the analysis by weighting each 
event $i$ with the beam polarisation value $P_{i}=P(T_{\rm n})$ 
at the measured neutron energy $T_{\rm n}$. The reconstruction 
efficiency $a_{{\rm p}{\rm p}\pi^{-}}$ shows a strong dependence 
on the kinematical variable $\theta_{\pi}^{*}$ and in particular 
on $M_{\rm pp}$\,\cite{DAU1,LAC1}. For very small $M_{\rm pp}$ 
values, both proton tracks are close together and the efficiency 
drops. As a consequence, each event was additionally weighted by 
the inverse of the reconstruction efficiency, 
$a_{{\rm p}{\rm p}\pi^{-}}^{-1}(M_{\rm pp},\theta_{\pi}^{*})$. 
This results in the following matrix equation for the estimators 
of the analysing powers $A_{N0}$ and $A_{S0}$:
\begin{eqnarray*}
\!\!\!\!\!\!\!\!\!\!\!\!\!\!\!\!\!\!\!\!\!\!\!\!\!\!\!\!\!\!\!\!\!\!\!\!\!\!\!\!\!\!\!\!\!\!\!\!\!\!\!\!\!\!\!\!\!\!\!\!\!\!\!\!\!\!\!\!\!\!\!\!\!\!\!\!\!\!\!\!\!\!\!\!\!\!\!\!\!\!\!\left(
\begin{array}{c}
\sum a_{i}^{-1} P_{i} \cos{\phi_{i}}\\
\sum a_{i}^{-1} P_{i} \sin{\phi_{i}}\\
\end{array}
\right)
=
\end{eqnarray*}
\begin{eqnarray}\label{weightedsums}
\!\!\!\!\left(
\begin{array}{ll}
\sum a_{i}^{-1} P_{i}^2 \cos^{2}{\phi_{i}}       & \sum a_{i}^{-1} P_{i}^2 \sin{\phi_{i}}\cos{\phi_{i}}\\
\sum a_{i}^{-1} P_{i}^2 \sin{\phi_{i}}\cos{\phi_{i}} & \sum a_{i}^{-1} P_{i}^2 \sin^{2}{\phi_{i}}
\end{array}
\right)\!\!
\left(
\begin{array}{c}
A_{N0}\\
A_{S0}
\end{array}
\right)
\end{eqnarray}
where the index $i$ runs over all events passing the 
reconstruction cuts.
}

\section{Results and discussion}

\subsection{Elastic scattering 
\boldmath $\vec{\rm n}{\rm p} \rightarrow {\rm n}{\rm p}$}

\begin{table*}
  \begin{center}
    \begin{tabular}{ccrccccrccccrcc}
    \hline
    \hline
$T_{\rm n}$ & $\theta_{\rm n}^{*}$ & $A_{00n0}$ & $\sigma_{\rm stat}$ & $\sigma_{\rm sys}$ &
$T_{\rm n}$ & $\theta_{\rm n}^{*}$ & $A_{00n0}$ & $\sigma_{\rm stat}$ & $\sigma_{\rm sys}$ &
$T_{\rm n}$ & $\theta_{\rm n}^{*}$ & $A_{00n0}$ & $\sigma_{\rm stat}$ & $\sigma_{\rm sys}$ \\
    \hline
    \hline
284         & 113.0                &   -0.108   &   0.111             & 0.005              &
404         & 112.9                &   -0.448   &   0.077             & 0.019              &
496         & 112.9                &   -0.321   &   0.075             & 0.013      \\   
            & 117.8                &   -0.194   &   0.070             & 0.009              &     
            & 117.8                &   -0.318   &   0.052             & 0.013              &
            & 117.8                &   -0.317   &   0.053             & 0.012      \\      
            & 122.8                &   -0.119   &   0.048             & 0.006              &  
            & 122.8                &   -0.125   &   0.037             & 0.005              &
            & 122.7                &   -0.234   &   0.039             & 0.009      \\    
            & 127.6                &   -0.125   &   0.039             & 0.006              &  
            & 127.5                &   -0.151   &   0.031             & 0.006              & 
            & 127.5                &   -0.171   &   0.035             & 0.007      \\       
            & 132.5                &   -0.106   &   0.037             & 0.005              &  
            & 132.5                &   -0.191   &   0.030             & 0.008              &  
            & 132.6                &   -0.142   &   0.033             & 0.006      \\   
            & 137.5                &   -0.137   &   0.036             & 0.006              &  
            & 137.5                &   -0.138   &   0.029             & 0.006              &
            & 137.5                &   -0.136   &   0.031             & 0.005      \\          
            & 142.5                &   -0.095   &   0.037             & 0.004              &  
            & 142.5                &   -0.116   &   0.028             & 0.005              & 
            & 142.5                &   -0.156   &   0.031             & 0.006      \\     
            & 147.5                &   -0.133   &   0.038             & 0.006              &  
            & 147.5                &   -0.074   &   0.029             & 0.003              & 
            & 147.5                &   -0.142   &   0.031             & 0.006      \\     
            & 152.5                &   -0.137   &   0.039             & 0.006              &   
            & 152.5                &   -0.108   &   0.030             & 0.005              & 
            & 152.5                &   -0.100   &   0.031             & 0.004      \\    
            & 157.4                &   -0.075   &   0.041             & 0.004              &   
            & 157.5                &   -0.091   &   0.032             & 0.004              &  
            & 157.4                &   -0.103   &   0.034             & 0.004      \\    
            & 162.4                &   -0.003   &   0.046             & 0.002              &   
            & 162.4                &   -0.069   &   0.037             & 0.003              &  
            & 162.3                &   -0.040   &   0.039             & 0.002      \\      
            & 167.1                &   -0.074   &   0.062             & 0.004              &   
            & 167.2                &   -0.033   &   0.050             & 0.001              &  
            & 167.1                &   -0.029   &   0.053             & 0.001      \\  
            & 171.9                &    0.041   &   0.112             & 0.002              &    
            & 171.8                &   -0.029   &   0.088             & 0.001              & 
            & 171.8                &   -0.094   &   0.095             & 0.004      \\  
            & 176.3                &   -0.260   &   0.370             & 0.012              &  
            & 176.4                &    0.041   &   0.277             & 0.017              &   
            & 176.3                &   -0.133   &   0.285             & 0.005      \\          
    \hline
314         & 113.0                &   -0.057   &   0.096             & 0.003              &
435         & 112.9                &   -0.178   &   0.072             & 0.007              &
525         & 112.8                &   -0.338   &   0.055             & 0.013      \\    
            & 117.8                &   -0.334   &   0.062             & 0.016              &
            & 117.8                &   -0.291   &   0.049             & 0.011              & 
            & 117.8                &   -0.354   &   0.038             & 0.013      \\   
            & 122.8                &   -0.104   &   0.042             & 0.005              &
            & 122.8                &   -0.224   &   0.035             & 0.008              & 
            & 122.7                &   -0.173   &   0.028             & 0.006      \\    
            & 127.6                &   -0.204   &   0.034             & 0.010              &   
            & 127.5                &   -0.206   &   0.031             & 0.008              &
            & 127.5                &   -0.234   &   0.025             & 0.009      \\  
            & 132.5                &   -0.137   &   0.033             & 0.006              &
            & 132.6                &   -0.142   &   0.029             & 0.005              &
            & 132.5                &   -0.148   &   0.024             & 0.006      \\  
            & 137.5                &   -0.149   &   0.032             & 0.007              &
            & 137.5                &   -0.126   &   0.028             & 0.005              &
            & 137.5                &   -0.145   &   0.023             & 0.005      \\   
            & 142.5                &   -0.104   &   0.032             & 0.005              & 
            & 142.5                &   -0.153   &   0.028             & 0.006              & 
            & 142.5                &   -0.111   &   0.022             & 0.004      \\ 
            & 147.5                &   -0.103   &   0.033             & 0.005              &
            & 147.5                &   -0.144   &   0.028             & 0.005              &
            & 147.5                &   -0.108   &   0.023             & 0.004      \\ 
            & 152.5                &   -0.071   &   0.034             & 0.003              &
            & 152.5                &   -0.150   &   0.029             & 0.006              &
            & 152.5                &   -0.055   &   0.023             & 0.002      \\ 
            & 157.4                &   -0.085   &   0.036             & 0.004              & 
            & 157.4                &   -0.081   &   0.031             & 0.003              &
            & 157.4                &   -0.103   &   0.024             & 0.004      \\ 
            & 162.4                &   -0.089   &   0.041             & 0.004              & 
            & 162.3                &   -0.019   &   0.036             & 0.001              &
            & 162.3                &   -0.069   &   0.029             & 0.003      \\    
            & 167.1                &   -0.093   &   0.055             & 0.004              & 
            & 167.1                &   -0.059   &   0.048             & 0.002              & 
            & 167.1                &   -0.075   &   0.039             & 0.003      \\   
            & 171.9                &    0.076   &   0.100             & 0.004              & 
            & 171.9                &   -0.218   &   0.084             & 0.008              &
            & 171.9                &   -0.010   &   0.069             & 0.000      \\  
            & 176.3                &   -0.110   &   0.309             & 0.005              & 
            & 177.3                &   -0.315   &   0.241             & 0.012              &
            & 176.3                &   -0.336   &   0.196             & 0.013      \\   
    \hline
344         & 113.0                &   -0.306   &   0.084             & 0.013              & 
465         & 112.9                &   -0.151   &   0.076             & 0.006              &    
550         & 112.9                &   -0.284   &   0.063             & 0.011      \\ 
            & 117.8                &   -0.207   &   0.056             & 0.009              &
            & 117.8                &   -0.289   &   0.052             & 0.011              &  
            & 117.8                &   -0.229   &   0.044             & 0.009      \\  
            & 122.8                &   -0.182   &   0.039             & 0.008              &
            & 122.7                &   -0.279   &   0.038             & 0.011              & 
            & 122.7                &   -0.194   &   0.034             & 0.007      \\    
            & 127.6                &   -0.184   &   0.032             & 0.008              &
            & 127.6                &   -0.189   &   0.033             & 0.007              &    
            & 127.6                &   -0.216   &   0.031             & 0.008      \\ 
            & 132.5                &   -0.146   &   0.031             & 0.006              &
            & 132.5                &   -0.166   &   0.031             & 0.006              &    
            & 132.6                &   -0.200   &   0.029             & 0.008      \\  
            & 137.5                &   -0.194   &   0.031             & 0.009              & 
            & 137.5                &   -0.107   &   0.030             & 0.004              & 
            & 137.5                &   -0.157   &   0.028             & 0.006      \\  
            & 142.5                &   -0.147   &   0.030             & 0.007              &
            & 142.5                &   -0.141   &   0.029             & 0.006              & 
            & 142.5                &   -0.103   &   0.027             & 0.004      \\  
            & 147.5                &   -0.161   &   0.031             & 0.007              &
            & 147.5                &   -0.137   &   0.030             & 0.005              &  
            & 147.5                &   -0.068   &   0.027             & 0.003      \\ 
            & 152.5                &   -0.033   &   0.032             & 0.002              &
            & 152.5                &   -0.079   &   0.031             & 0.003              & 
            & 152.5                &   -0.087   &   0.028             & 0.003      \\ 
            & 157.4                &   -0.172   &   0.034             & 0.008              &
            & 157.4                &   -0.096   &   0.033             & 0.004              &  
            & 157.4                &   -0.099   &   0.030             & 0.004      \\    
            & 162.3                &   -0.090   &   0.038             & 0.004              &
            & 162.3                &   -0.088   &   0.038             & 0.003              &  
            & 162.3                &   -0.094   &   0.034             & 0.004      \\  
            & 167.1                &   -0.033   &   0.052             & 0.002              &
            & 167.1                &   -0.005   &   0.052             & 0.000              &  
            & 167.1                &    0.013   &   0.048             & 0.001      \\    
            & 171.8                &    0.020   &   0.094             & 0.001              & 
            & 171.8                &   -0.170   &   0.091             & 0.007              & 
            & 171.9                &   -0.104   &   0.086             & 0.004      \\  
            & 176.3                &    0.227   &   0.292             & 0.010              & 
            & 176.5                &   -0.334   &   0.293             & 0.013              &    
            & 176.3                &   -0.333   &   0.243             & 0.013      \\   
    \hline
374         & 112.9                &   -0.237   &   0.081             & 0.010              & 
            &                      &            &                     &                    &
            &                      &            &                     &            \\
            & 117.8                &   -0.355   &   0.054             & 0.016              &
            &                      &            &                     &                    &
            &                      &            &                     &            \\
            & 122.8                &   -0.199   &   0.038             & 0.009              & 
            &                      &            &                     &                    &
            &                      &            &                     &            \\
            & 127.6                &   -0.141   &   0.032             & 0.006              & 
            &                      &            &                     &                    &
            &                      &            &                     &            \\
            & 132.5                &   -0.123   &   0.030             & 0.005              & 
            &                      &            &                     &                    &
            &                      &            &                     &            \\
            & 137.5                &   -0.176   &   0.030             & 0.008              & 
            &                      &            &                     &                    &
            &                      &            &                     &            \\
            & 142.5                &   -0.124   &   0.029             & 0.005              &
            &                      &            &                     &                    &
            &                      &            &                     &            \\
            & 147.5                &   -0.104   &   0.030             & 0.005              & 
            &                      &            &                     &                    &
            &                      &            &                     &            \\
            & 152.5                &   -0.115   &   0.031             & 0.005              &
            &                      &            &                     &                    &
            &                      &            &                     &            \\
            & 157.5                &   -0.128   &   0.032             & 0.006              &
            &                      &            &                     &                    &
            &                      &            &                     &            \\
            & 162.5                &   -0.063   &   0.037             & 0.003              & 
            &                      &            &                     &                    &
            &                      &            &                     &            \\
            & 167.5                &   -0.101   &   0.051             & 0.004              & 
            &                      &            &                     &                    &
            &                      &            &                     &            \\
            & 172.5                &    0.039   &   0.090             & 0.002              &
            &                      &            &                     &                    &
            &                      &            &                     &            \\
            & 177.5                &    0.644   &   0.286             & 0.028              & 
            &                      &            &                     &                    &
            &                      &            &                     &            \\  
    \hline
    \end{tabular}
  \caption{\label{a00n0results} 
   Analysing powers $A_{00n0}$ for the np elastic scattering. 
   Quoted are statistical and systematic errors.}
  \end{center}
\end{table*}

\noindent{
As a by-product, analysing powers for the np elastic scattering,
using events from the minimum bias trigger sample, have been 
measured. Adopting the convention of Ref.\,\cite{BYS2}, the 
analysing powers of interest for elastic scattering are denoted 
$A_{00n0}$ and $A_{00s0}$. They correspond to $A_{N0}$ and $A_{S0}$ 
by replacing $\vec{p_{\pi}^{*}}$ with the momentum vector of the 
scattered neutron $\vec{p_{\rm n}^{*\prime}}$. 
For elastic scattering, there is only one independent kinematic 
variable in the final state. As a consequence, the weighting of 
the events by the reconstruction efficiency was omitted in 
(\ref{weightedsums}).
}

\begin{figure*}
\resizebox{1.0\textwidth}{!}{%
  \includegraphics{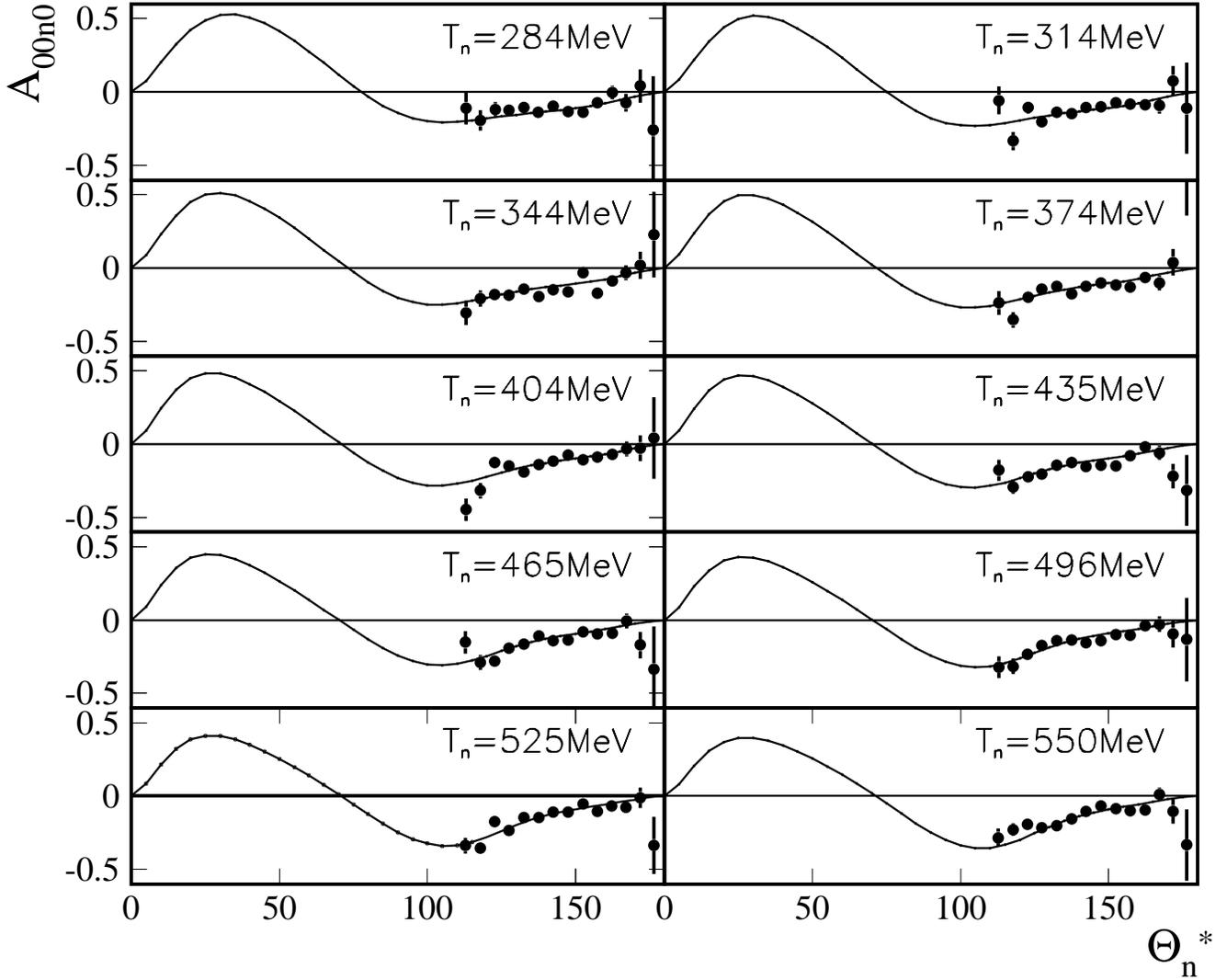}
}
\vspace{-0.0cm}       
\caption{
  np elastic scattering: analysing powers $A_{00n0}$ as a function of the 
  neutron c.m. scattering angle $\theta_{\rm n}^{*}$ (full dots). Shown are 
  statistical errors only. Solid line: results from the partial wave analysis 
  of Arndt et al.\,\cite{ARND1} where the new data have not been included.
}
\label{a00n0}      
\end{figure*}

\noindent{
For the determination of the analysing powers, the data with
neutron energies between 270 and 570 MeV were subdivided in 
10 bins of equal width. For each neutron energy bin, the mean 
neutron energy was computed. The analysing powers $A_{00n0}$ 
and $A_{00s0}$ were calculated as a function of the neutron 
c.m. scattering angle $\theta_{\rm n}^{*}$. The results for 
$A_{00n0}$ are shown in Fig.\,\ref{a00n0}. The numerical values 
are given in Tab.\,\ref{a00n0results}.
}

\noindent{
A contribution to the systematic uncertainty is the error on the 
neutron beam polarisation of about 3~\%. Background contributions 
from the target surroundings and drift chamber materials were 
determined from runs with an empty target cell and found to be 
12~\% averaged over the considered neutron energies. The spin 
dependent asymmetry from this background source showed similar 
results as the data with the full target cell. However, these 
asymmetries were determined with much less statistical precision. 
Under the assumption that the asymmetry from this background 
differs by $\pm 20~\%$ from the asymmetry of signal events, the 
relative systematic error was estimated to be $\pm 2~\%$. Inelastic 
reactions in the liquid hydrogen target gave only a small contribution 
of $3~\%$ at 270 MeV and $1~\%$ at 550 MeV the asymmetry of which 
could not be determined. Under the conservative assumption that the 
analysing power of this background can take any value between +1 and 
$-$1, an additional systematic error was assigned which reads $\pm 3~\%$ 
at 270 MeV and $1~\%$ at 550 MeV. All systematic error contributions  
were added in quadrature.
}

\noindent{
As can be seen from Fig.\,\ref{a00n0}, the analysing powers $A_{00n0}$ 
are in good agreement with the results of a partial wave analysis 
performed by Arndt et al.\,\cite{ARND1} where the new data have not been 
taken into account. The analysing powers $A_{00s0}$ for the different 
neutron energy bins are consistent with zero as it is expected due to 
parity conservation\,\cite{LAC1}. 
}

\noindent{
Under the assumption of parity conservation, the ratio 
$A_{00s0}/A_{00n0}$ allows to test if the horizontally 
(vertically) polarised beam contained additional, small 
polarisation components in the vertical (horizontal) 
direction. For all neutron energies, this ratio was 
found to be consistent with zero within the statistical 
uncertainties. Averaging over all neutron energies, an 
asymmetry
\begin{equation*} 
<\! \epsilon_{s}\!>=<\!P\!>\cdot<\!A_{00s0}\!>= -0.0002 \pm 0.001
\end{equation*} 
was found. This has to be compared to the energy averaged 
asymmetry 
\begin{equation*} 
<\!\epsilon_{n}\!> = <\!P\!> \cdot <\!A_{00n0}\!> = 0.05 \pm 0.001.
\end{equation*} 
Hence, the relative contribution of a transverse component 
perpendicular to the main transverse polarisation was 
smaller than 3\% at 68\% confidence level. This finding
is in agreement with the asymmetries determined from the 
beam monitor scaler rates.
}

\subsection{Analysing powers for 
\boldmath $\vec{\rm n}{\rm p} \rightarrow {\rm p}{\rm p} \pi^{-}$}

\noindent{
For the determination of the analysing powers for the reaction 
$\vec{\rm n}{\rm p} \rightarrow {\rm p}{\rm p} \pi^{-}$, the data 
were subdivided in nine neutron energy bins where the first bin 
was between threshold and 330 MeV while the other bins were of 
30 MeV width. For each neutron energy bin, the mean neutron energy 
was calculated from the neutron energy distribution. The results 
are presented as a function of $\cos{\theta_{\pi}^{*}}$.
}

\noindent{
In general, statistical errors are the main uncertainties. With 
increasing neutron energy, the statistical error decreases and 
the systematic error becomes more and more important. Above 465
MeV, they even surpass the statistical error in certain regions 
of $\theta_{\pi}^{*}$.
The systematic error contains various contributions:
\begin{enumerate}
\item An uncertainty of $\pm 3~\%$ due to the experimental 
      error in the beam polarisation.
\item The asymmetry from background events produced in the 
      target surroundings. Its effect was determined using 
      data taken with an empty target cell. However, the 
      statistical precision was significantly smaller than 
      for the data with the full target cell. For energies 
      above 400 MeV, the asymmetries from this background 
      were found to have the same sign as the asymmetries 
      with the full target cell. However, they were smaller 
      in magnitude by about a factor of two. According to 
      the background contribution, between $4~\%$ at 550 MeV 
      and $6~\%$ at 400 MeV, the asymmetries were enlarged 
      in magnitude by $2~\%$ to $3~\%$, respectively.
      For energies below 400 MeV, the background asymmetries 
      were consistent with zero on average. The background 
      contribution in the data with full target cell increases
      with decreasing energy and reads $8~\%$ at 315 MeV. As 
      a consequence, the asymmetry was corrected by the same 
      size. Since the statistical error for the empty target
      measurement is quite large, an additional systematic 
      error of the size of the correction was assigned.
\item If the velocities of the three emitted particles are 
      similar, the kinematical fit procedure possibly assigns 
      the wrong particle hypothesis to the measured tracks\,\cite{DAU1}. 
      From a conservative estimate, using the detector Monte 
      Carlo simulation, it was concluded that this happens in 
      less than $5~\%$ of the events. This effect could lead 
      to a bias by reducing the measured asymmetry. It was 
      taken into account by increasing the value of $A_{N0}$ 
      by $\pm 2.5~\%$ and assigning an additional systematic 
      error of the same size.
\end{enumerate}
The various systematic errors have been added in quadrature.
}

\begin{figure*}
\resizebox{1.0\textwidth}{!}{%
  \includegraphics{fig.2}
}
\vspace{-0.0cm}       
\caption{
  Analysing powers $A_{N0}$ for the nine bins in neutron energy 
  as a function of $\cos{\theta_{\pi}^{*}}$ ($\bullet$). The 
  statistical errors are indicated by the error bars. The 
  systematic error is indicated by the shaded band.
  Also shown are data for the reaction 
  $\vec{\rm p}{\rm p} \rightarrow {\rm p}{\rm p} \pi^{0}$ 
  from two experiments: for proton energies $T_{\rm p}=$ 325, 460, 
  500, 520 and 540 MeV from Ref.\,\cite{RAP1}($\square$) 
  and for proton energies 
  $T_{\rm p}=$ 325, 350, 375, 400 MeV from Ref.\,\cite{MEY3}($\circ$).
}
\label{an0}      
\end{figure*}

\noindent{
The numerical results for $A_{N0}(\cos{\theta_{\pi}^{*}})$ are presented 
in Tab.\,\ref{an0results}. In general, large negative values for the 
analysing powers $A_{N0}(\cos{\theta_{\pi}^{*}})$ are observed, as can be 
seen in Fig.\,\ref{an0}. At 315 MeV, values compatible with zero are found 
in the forward and backward direction, whereas negative values are observed 
around $\cos{\theta_{\pi}^{*}} \approx 0$. At 345 MeV, a positive value, 
though consistent with zero, is found in the backward direction. 
At intermediate energies, the angular dependence of $A_{N0}$ is more or 
less forward-backward symmetric whereas at higher energies a significant 
forward-backward asymmetry is observed.
}

\begin{figure}
\resizebox{0.5\textwidth}{!}{%
  \includegraphics{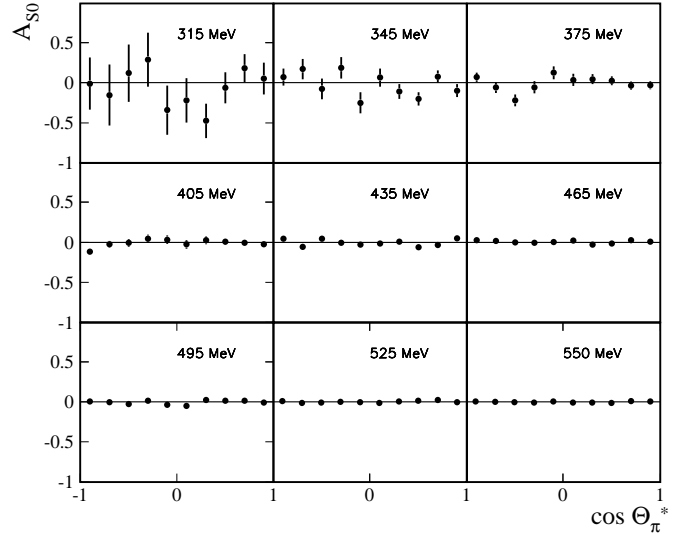}
}
\vspace{-0.0cm}       
\caption{
  Analysing powers $A_{S0}$ as a function $\cos{\theta_{\pi}^{*}}$. 
}
\label{as0}      
\end{figure}

\noindent{
The results for the analysing powers $A_{S0}(\cos{\theta_{\pi}^{*}})$ 
are shown in Fig.\,\ref{as0}. Averaging the 
$A_{S0}(\cos{\theta_{\pi}^{*}})$ over $\cos\theta_{\pi}^{*}$, 
the maximal deviation from zero is found to be 1.1 standard 
deviations. The asymmetries, averaged over $\theta_{\pi}^{*}$, 
show the smallest statistical errors at $T_{\rm n}=525~{\rm MeV}$. 
They read
\begin{equation*} 
<\! \epsilon_{S}\!>=<\!P\!>\cdot<\!A_{S0}\!>= 0.0007 \pm 0.0014
\end{equation*} 
and
\begin{equation*} 
<\!\epsilon_{N}\!> = <\!P\!> \cdot <\!A_{N0}\!> = -0.1024 \pm 0.0014.
\end{equation*} 
Hence, the $A_{S0}(\cos{\theta_{\pi}^{*}})$-values are consistent 
with zero, which is in agreement with the result observed in the 
elastic np scattering case. As a consequence, also for the 
three-body final state, no significant bias from detector asymmetries 
or beam properties is observed within the available statistical 
precision.
}

\subsection{Comparison with other experiments}

\subsubsection{Proton-Proton experiments}

\noindent{
In an experiment described in Ref.\,\cite{STA1}, analysing powers 
for the reaction 
$\vec{\rm p}{\rm p} \rightarrow {\rm p}{\rm p} \pi^{0}$ have been 
measured and found to be negative for all beam energies between 
319 MeV and 496 MeV. However, these results were presented in the 
laboratory system only. Due to the different experimental set-ups, 
they can not be directly compared to our data. 
}

\noindent{
In a SATURNE experiment, analysing powers from the reaction
$\vec{\rm p}{\rm p} \rightarrow {\rm p}{\rm p} \pi^{0}$
have been measured at various proton beam energies between 
325 MeV and 1012 MeV\,\cite{RAP1}. 
}

\noindent{
Although these data do not cover the full angular range for 
all beam energies, they suggest to be forward-backward 
symmetric. The negative values, observed for energies above 
460 MeV, were interpreted as an interference between Ps and 
Pp partial waves from $\sigma_{11}$\,\cite{RAP1}. 
}

\noindent{
For high energies (above 460 MeV), the asymmetries of 
Ref.\,\cite{RAP1}, shown in Fig.\,\ref{an0} as boxes, 
differ in the forward direction 
($\cos{\theta_{\pi}^{*}} \approx 0.5$) in a significant 
way from those measured in 
$\vec{\rm n}{\rm p} \rightarrow {\rm p}{\rm p} \pi^{-}$.
This is a clear signal that $\sigma_{01}$ is present in 
the reaction $\vec{\rm n}{\rm p} \rightarrow {\rm p}{\rm p} \pi^{-}$.
The difference is observed at high energies where the pion 
production mechanism for $\sigma_{11}$ is already dominated 
by the excitation of an intermediate $N\Delta$ state. 
Hence, $\sigma_{01}$ is still of importance for
$\vec{\rm n}{\rm p} \rightarrow {\rm p}{\rm p} \pi^{-}$
even at energies where resonant pion production dominates. 
This qualitative finding is in agreement with the result 
from ref.~\cite{DAU1} where in the same energy region a 
$20-30~\%$ contribution of $\sigma_{01}$ to the total 
$\vec{\rm n}{\rm p} \rightarrow {\rm p}{\rm p} \pi^{-}$
cross section has been reported.
}

\noindent{
The asymmetries from Ref.\,\cite{RAP1} at $T_{\rm p}=325~{\rm MeV}$ 
are slightly positive and differ at $\cos{\theta_{\pi}^{*}} \approx 0$ 
from our $\vec{\rm n}{\rm p} \rightarrow {\rm p}{\rm p} \pi^{-}$
results though our uncertainties are large. Measurements performed 
at the Indiana Cooler synchrotron gave negative asymmetries in the 
reaction $\vec{\rm p}{\rm p} \rightarrow {\rm p}{\rm p} \pi^{0}$ for 
all proton beam energies between $T_{\rm p}=325~{\rm MeV}$ and 400 
MeV\,\cite{MEY3}. Their results are shown in Fig.\,\ref{an0} as open 
cirles. Again, for $T_{\rm p}=325~{\rm MeV}$ , their results differ 
from $\vec{\rm n}{\rm p} \rightarrow {\rm p}{\rm p} \pi^{-}$ around 
$\cos{\theta_{\pi}^{*}} \approx 0$. For proton energies at 350 MeV 
and 375 MeV, the statistical accuracy in the 
$\vec{\rm n}{\rm p} \rightarrow {\rm p}{\rm p} \pi^{-}$ results still 
does not allow to state significant differences between both reactions. 
However, at 400 MeV the analysing power reported in Ref.\,\cite{MEY3} 
clearly differs from the 
$\vec{\rm n}{\rm p} \rightarrow {\rm p}{\rm p} \pi^{-}$ data.
}

\begin{table*}
  \begin{center}
    \begin{tabular}{crrrrrcrrrrr}
    \hline
    \hline
$T_{\rm n}$ (MeV) & $\cos{\theta_{\pi}^{*}}$ & $A_{N0}$ & $\sigma_{\rm stat}$ & $\sigma_{\rm sys}$ & events &
$T_{\rm n}$ (MeV) & $\cos{\theta_{\pi}^{*}}$ & $A_{N0}$ & $\sigma_{\rm stat}$ & $\sigma_{\rm sys}$ & events\\
    \hline
    \hline
315 & -0.9   &   -0.212  &   0.323   & 0.017   &      262  & 465 & -0.9   &   -0.217  &   0.019  & 0.010    &   36673   \\    
    & -0.7   &   -0.255  &   0.379   & 0.021   &      190  &     & -0.7   &   -0.370  &   0.022  & 0.017    &   28701   \\    
    & -0.5   &   -0.427  &   0.357   & 0.034   &      213  &     & -0.5   &   -0.472  &   0.024  & 0.022    &   24202   \\    
    & -0.3   &   -0.214  &   0.338   & 0.017   &      240  &     & -0.3   &   -0.475  &   0.026  & 0.022    &   20316   \\    
    & -0.1   &   -1.088  &   0.299   & 0.097   &      293  &     & -0.1   &   -0.411  &   0.028  & 0.019    &   17192   \\    
    &  0.1   &   -0.910  &   0.271   & 0.073   &      359  &     &  0.1   &   -0.383  &   0.029  & 0.018    &   16083   \\    
    &  0.3   &    0.218  &   0.214   & 0.018   &      597  &     &  0.3   &   -0.325  &   0.027  & 0.015    &   19361   \\    
    &  0.5   &   -0.378  &   0.194   & 0.030   &      723  &     &  0.5   &   -0.248  &   0.024  & 0.012    &   23348   \\    
    &  0.7   &    0.043  &   0.176   & 0.003   &      887  &     &  0.7   &   -0.171  &   0.023  & 0.008    &   26464   \\    
    &  0.9   &    0.427  &   0.197   & 0.034   &      700  &     &  0.9   &   -0.029  &   0.022  & 0.001    &   27803   \\    
    \hline
345 & -0.9   &    0.188  &   0.106   & 0.014   &     1735  & 495 & -0.9   &   -0.217  &   0.015  & 0.009    &   76544   \\    
    & -0.7   &   -0.037  &   0.125   & 0.003   &     1246  &     & -0.7   &   -0.382  &   0.017  & 0.016    &   59106   \\    
    & -0.5   &   -0.342  &   0.129   & 0.025   &     1165  &     & -0.5   &   -0.412  &   0.019  & 0.017    &   49809   \\    
    & -0.3   &   -0.374  &   0.135   & 0.027   &     1065  &     & -0.3   &   -0.437  &   0.020  & 0.018    &   41635   \\    
    & -0.1   &   -0.405  &   0.131   & 0.030   &     1119  &     & -0.1   &   -0.429  &   0.022  & 0.018    &   34785   \\    
    &  0.1   &   -0.230  &   0.115   & 0.017   &     1485  &     &  0.1   &   -0.362  &   0.023  & 0.015    &   31131   \\    
    &  0.3   &   -0.239  &   0.093   & 0.018   &     2227  &     &  0.3   &   -0.271  &   0.022  & 0.011    &   37177   \\    
    &  0.5   &   -0.244  &   0.084   & 0.018   &     2790  &     &  0.5   &   -0.194  &   0.020  & 0.008    &   43431   \\    
    &  0.7   &   -0.189  &   0.075   & 0.014   &     3471  &     &  0.7   &   -0.148  &   0.018  & 0.006    &   50670   \\    
    &  0.9   &   -0.050  &   0.080   & 0.004   &     3072  &     &  0.9   &   -0.031  &   0.018  & 0.001    &   54861   \\    
    \hline
375 & -0.9   &   -0.051  &   0.058   & 0.003   &     5195  & 525 & -0.9   &   -0.232  &   0.008  & 0.010    &  178779  \\    
    & -0.7   &   -0.189  &   0.068   & 0.013   &     3785  &     & -0.7   &   -0.407  &   0.010  & 0.017    &  135787  \\    
    & -0.5   &   -0.264  &   0.074   & 0.017   &     3175  &     & -0.5   &   -0.439  &   0.012  & 0.018    &  114000  \\    
    & -0.3   &   -0.261  &   0.077   & 0.017   &     2937  &     & -0.3   &   -0.423  &   0.012  & 0.018    &   93306  \\    
    & -0.1   &   -0.425  &   0.079   & 0.028   &     2716  &     & -0.1   &   -0.359  &   0.013  & 0.015    &   75911  \\    
    &  0.1   &   -0.378  &   0.075   & 0.025   &     3075  &     &  0.1   &   -0.303  &   0.014  & 0.013    &   66249  \\    
    &  0.3   &   -0.493  &   0.063   & 0.033   &     4261  &     &  0.3   &   -0.222  &   0.013  & 0.009    &   77681  \\    
    &  0.5   &   -0.262  &   0.056   & 0.017   &     5568  &     &  0.5   &   -0.174  &   0.012  & 0.007    &   90419  \\    
    &  0.7   &   -0.115  &   0.051   & 0.008   &     6595  &     &  0.7   &   -0.106  &   0.011  & 0.004    &  105369  \\    
    &  0.9   &   -0.100  &   0.052   & 0.007   &     6335  &     &  0.9   &   -0.044  &   0.010  & 0.002    &  116920  \\    
    \hline
405 & -0.9   &   -0.113  &   0.039   & 0.005  &     10287  & 550 & -0.9   &   -0.270  &   0.008  & 0.011    &  157751  \\    
    & -0.7   &   -0.217  &   0.044   & 0.010  &      7954  &     & -0.7   &   -0.416  &   0.010  & 0.017    &  120750  \\    
    & -0.5   &   -0.454  &   0.047   & 0.021  &      6743  &     & -0.5   &   -0.448  &   0.010  & 0.019    &  101710  \\    
    & -0.3   &   -0.437  &   0.051   & 0.020  &      5888  &     & -0.3   &   -0.438  &   0.012  & 0.018    &   82258  \\    
    & -0.1   &   -0.480  &   0.053   & 0.022  &      5366  &     & -0.1   &   -0.381  &   0.013  & 0.016    &   66408  \\    
    &  0.1   &   -0.464  &   0.053   & 0.022  &      5399  &     &  0.1   &   -0.315  &   0.014  & 0.013    &   57700  \\    
    &  0.3   &   -0.360  &   0.047   & 0.017  &      6941  &     &  0.3   &   -0.209  &   0.013  & 0.009    &   67116  \\    
    &  0.5   &   -0.273  &   0.041   & 0.013  &      8939  &     &  0.5   &   -0.140  &   0.012  & 0.006    &   77921  \\    
    &  0.7   &   -0.231  &   0.039   & 0.011  &     10263  &     &  0.7   &   -0.060  &   0.011  & 0.003    &   89622  \\    
    &  0.9   &   -0.069  &   0.039   & 0.003  &     10196  &     &  0.9   &   -0.021  &   0.010  & 0.001    &  105223  \\    
    \hline
435 & -0.9   &   -0.126  &   0.025   & 0.006  &     19302  &     &        &           &          &     &   \\    
    & -0.7   &   -0.355  &   0.029   & 0.017  &     14855  &     &        &           &          &     &   \\    
    & -0.5   &   -0.341  &   0.031   & 0.016  &     12846  &     &        &           &          &     &   \\    
    & -0.3   &   -0.488  &   0.033   & 0.023  &     11090  &     &        &           &          &     &   \\    
    & -0.1   &   -0.462  &   0.036   & 0.022  &      9484  &     &        &           &          &     &   \\   
    &  0.1   &   -0.418  &   0.036   & 0.019  &      9232  &     &        &           &          &     &   \\   
    &  0.3   &   -0.302  &   0.033   & 0.014  &     11421  &     &        &           &          &     &   \\    
    &  0.5   &   -0.182  &   0.029   & 0.009  &     13841  &     &        &           &          &     &   \\    
    &  0.7   &   -0.095  &   0.028   & 0.004  &     15962  &     &        &           &          &     &   \\    
    &  0.9   &   -0.086  &   0.028   & 0.004  &     16256  &     &        &           &          &     &   \\    
    \hline
    \end{tabular}
  \caption{\label{an0results} 
   Analysing powers $A_{N0}$ for the reaction 
   ${\rm n}{\rm p} \rightarrow {\rm p}{\rm p} \pi^{-}$.
   Quoted are statistical and systematic errors.}
  \end{center}
\end{table*}

\subsubsection{Neutron-Proton experiments}
\noindent{
In a TRIUMF experiment, analysing powers 
$A_{N0}(\cos{\theta_{\pi}^{*}})$ were measured\,\cite{BAC1} 
at 443 MeV and presented in different bins of $M_{\rm pp}$. 
At 435 MeV, we calculated $A_{N0}(\cos{\theta_{\pi}^{*}})$ 
for the same $M_{\rm pp}$ binning as in Ref.\,\cite{BAC1}. 
The numerical values are given in Tab.\,\ref{an0bachmanresults}. 
Overall, both data sets are in good agreement as can be seen 
from Fig.\,\ref{bachman}. In general, the analysing powers 
are negative with a slight forward-backward asymmetry. For 
the smallest $M_{\rm pp}$ bin, both experiments observe a 
zero-crossing in the backward direction. This finding was 
interpreted as the sign of an interference between Ss and 
Sp partial waves\,\cite{BAC1} and hence as an indication 
for $\sigma_{01}$. This interpretation was confirmed by 
the results of the TRIUMF experiments described in 
refs.\,\cite{DUNC1,DUNC2}.
}

\begin{figure}
\resizebox{0.5\textwidth}{!}{%
  \includegraphics{fig.4}
}
\vspace{-0.0cm}       
\caption{
  Comparison of $A_{N0}(\cos{\theta_{\pi}^{*}})$ values ($\bullet$) 
  at 435 MeV subdivided in different bins of $M_{\rm pp}$ with 
  the results of Ref.\,\cite{BAC1} at 443 MeV ($\circ$).
}
\label{bachman}      
\end{figure}

\noindent{
In a SATURNE experiment\,\cite{TER1}, analysing powers $A_{N0}(\theta_{\pi}^{*})$ 
were measured in different bins of $M_{\rm pp}$ at several neutron energies. The 
lowest neutron beam energy which can be compared with our results was at 572 MeV. 
Fig.\,\ref{terrien} shows their results as a function of $\cos\theta_{\pi}^{*}$ 
together with our $A_{N0}(\cos\theta_{\pi}^{*})$ values at 550 MeV using the same 
$M_{\rm pp}$ binning. The numerical values are given in Tab.\,\ref{an0terrienresults}.
Both experiments are in qualitative agreement. Quantitative deviations might be 
assigned to the difference in the beam energies. In both cases, the analysing 
powers are mainly negative. Compared to the results at 435 MeV, see Fig.\,\ref{bachman}, 
the forward-backward asymmetry is even more pronounced. Again, for the smallest 
$M_{\rm pp}$ bin, a zero-crossing is observed; however, this time in the forward 
direction. 
}

\begin{table*}
  \begin{center}
    \begin{tabular}{crrrrrcrrrrr}
    \hline
    \hline
$M_{\rm pp}$ (MeV) & $\cos{\theta_{\pi}^{*}}$ & $A_{N0}$ & $\sigma_{\rm stat}$ & $\sigma_{\rm sys}$ & events &
$M_{\rm pp}$ (MeV) & $\cos{\theta_{\pi}^{*}}$ & $A_{N0}$ & $\sigma_{\rm stat}$ & $\sigma_{\rm sys}$ & events\\
    \hline
    \hline
1876-1888          & -0.9                     &  0.361   & 0.084               & 0.018              & 1749    &
1912-1924          & -0.9                     & -0.239   & 0.053               & 0.012              & 4511   \\    
                   & -0.7                     & -0.092   & 0.116               & 0.005              &  934    &
                   & -0.7                     & -0.388   & 0.057               & 0.019              & 3845    \\    
                   & -0.5                     & -0.075   & 0.137               & 0.004              &  668    &
                   & -0.5                     & -0.338   & 0.061               & 0.017              & 3358    \\    
                   & -0.3                     & -0.031   & 0.146               & 0.002              &  589    &
                   & -0.3                     & -0.478   & 0.067               & 0.024              & 2729     \\    
                   & -0.1                     & -0.446   & 0.136               & 0.022              &  663    &
                   & -0.1                     & -0.442   & 0.077               & 0.022              & 2060     \\    
                   &  0.1                     & -0.495   & 0.125               & 0.024              &  785    & 
                   &  0.1                     & -0.471   & 0.077               & 0.023              & 2041      \\    
                   &  0.3                     & -0.073   & 0.095               & 0.004              & 1378    &
                   &  0.3                     & -0.330   & 0.073               & 0.016              & 2330     \\    
                   &  0.5                     & -0.074   & 0.072               & 0.004              & 2386    & 
                   &  0.5                     & -0.217   & 0.073               & 0.011              & 2359     \\    
                   &  0.7                     &  0.083   & 0.063               & 0.004              & 3197    &
                   &  0.7                     & -0.210   & 0.071               & 0.010              & 2487     \\    
                   &  0.9                     &  0.004   & 0.064               & 0.000              & 3033    &
                   &  0.9                     & -0.144   & 0.069               & 0.007              & 2606     \\    
    \hline
1888-1900          & -0.9                     & -0.193   & 0.050               & 0.009              & 4945    &
1924-1936          & -0.9                     & -0.280   & 0.076               & 0.014              & 2150   \\    
                   & -0.7                     & -0.299   & 0.060               & 0.015              & 3458    &
                   & -0.7                     & -0.531   & 0.084               & 0.026              & 1724    \\    
                   & -0.5                     & -0.379   & 0.064               & 0.019              & 3034    &
                   & -0.5                     & -0.343   & 0.094               & 0.017              & 1383    \\    
                   & -0.3                     & -0.515   & 0.067               & 0.025              & 2672    &
                   & -0.3                     & -0.693   & 0.107               & 0.034              & 1038     \\    
                   & -0.1                     & -0.444   & 0.070               & 0.022              & 2470    &
                   & -0.1                     & -0.572   & 0.118               & 0.028              &  873     \\    
                   &  0.1                     & -0.369   & 0.073               & 0.018              & 2287    & 
                   &  0.1                     & -0.397   & 0.114               & 0.020              &  947      \\    
                   &  0.3                     & -0.430   & 0.064               & 0.021              & 3024    &
                   &  0.3                     & -0.169   & 0.113               & 0.008              &  974     \\    
                   &  0.5                     & -0.126   & 0.056               & 0.006              & 4052    & 
                   &  0.5                     & -0.393   & 0.115               & 0.019              &  922     \\    
                   &  0.7                     & -0.123   & 0.051               & 0.006              & 4861    &
                   &  0.7                     & -0.196   & 0.109               & 0.010              & 1041     \\    
                   &  0.9                     & -0.020   & 0.051               & 0.001              & 4841    &
                   &  0.9                     & -0.208   & 0.100               & 0.010              & 1248     \\    
    \hline
1900-1912          & -0.9                     & -0.312   & 0.047               & 0.015              & 5603    &
1936-1948          & -0.9                     & -0.070   & 0.192               & 0.003              &  338   \\    
                   & -0.7                     & -0.425   & 0.051               & 0.021              & 4663    &
                   & -0.7                     & -0.422   & 0.232               & 0.021              &  227    \\    
                   & -0.5                     & -0.408   & 0.054               & 0.020              & 4211    &
                   & -0.5                     & -0.597   & 0.252               & 0.029              &  189    \\    
                   & -0.3                     & -0.501   & 0.056               & 0.025              & 3862    &
                   & -0.3                     & -0.514   & 0.248               & 0.025              &  198     \\    
                   & -0.1                     & -0.534   & 0.062               & 0.026              & 3193    &
                   & -0.1                     & -0.083   & 0.240               & 0.004              &  216     \\    
                   &  0.1                     & -0.436   & 0.064               & 0.022              & 2979    & 
                   &  0.1                     & -0.162   & 0.258               & 0.008              &  187      \\    
                   &  0.3                     & -0.406   & 0.059               & 0.020              & 3511    &
                   &  0.3                     & -0.229   & 0.252               & 0.011              &  196     \\    
                   &  0.5                     & -0.196   & 0.056               & 0.010              & 3911    & 
                   &  0.5                     & -0.175   & 0.245               & 0.009              &  207     \\    
                   &  0.7                     & -0.092   & 0.055               & 0.005              & 4141    &
                   &  0.7                     & -0.068   & 0.232               & 0.003              &  232     \\    
                   &  0.9                     & -0.167   & 0.054               & 0.008              & 4255    &
                   &  0.9                     & -0.053   & 0.214               & 0.003              &  272     \\    
    \hline
    \end{tabular}
  \caption{\label{an0bachmanresults} 
   Analysing powers $A_{N0}$ for the reaction ${\rm n}{\rm p} \rightarrow {\rm p}{\rm p} \pi^{-}$
   at $T_{\rm n}=435~{\rm MeV}$ for different bins in $M_{\rm pp}$.
   Quoted are statistical and systematic errors.}
  \end{center}
\end{table*}
\begin{table*}
  \begin{center}
    \begin{tabular}{crrrrrcrrrrr}
    \hline
    \hline
$M_{\rm pp}$ (MeV) & $\cos{\theta_{\pi}^{*}}$ & $A_{N0}$ & $\sigma_{\rm stat}$ & $\sigma_{\rm sys}$ & events &
$M_{\rm pp}$ (MeV) & $\cos{\theta_{\pi}^{*}}$ & $A_{N0}$ & $\sigma_{\rm stat}$ & $\sigma_{\rm sys}$ & events\\
    \hline
    \hline
1876-1902          & -0.9                     & -0.109   & 0.023               & 0.005              & 24525    &
1952-1977          & -0.9                     & -0.301   & 0.027               & 0.013              & 16778   \\    
                   & -0.7                     & -0.222   & 0.028               & 0.010              & 16291    &
                   & -0.7                     & -0.446   & 0.030               & 0.020              & 13970    \\    
                   & -0.5                     & -0.227   & 0.030               & 0.010              & 13512    &
                   & -0.5                     & -0.480   & 0.034               & 0.021              & 10644    \\    
                   & -0.3                     & -0.309   & 0.021               & 0.014              & 11954    &
                   & -0.3                     & -0.425   & 0.041               & 0.019              &  7209     \\    
                   & -0.1                     & -0.261   & 0.034               & 0.011              & 10565    &
                   & -0.1                     & -0.411   & 0.047               & 0.018              &  5544     \\    
                   &  0.1                     & -0.228   & 0.036               & 0.010              &  9597    & 
                   &  0.1                     & -0.497   & 0.046               & 0.022              &  5676      \\    
                   &  0.3                     & -0.094   & 0.031               & 0.004              & 13071    &
                   &  0.3                     & -0.223   & 0.044               & 0.010              &  6290     \\    
                   &  0.5                     & -0.051   & 0.026               & 0.002              & 18860    & 
                   &  0.5                     & -0.223   & 0.045               & 0.010              &  6137     \\    
                   &  0.7                     &  0.099   & 0.022               & 0.004              & 25013    &
                   &  0.7                     & -0.124   & 0.043               & 0.005              &  6809     \\    
                   &  0.9                     &  0.040   & 0.021               & 0.002              & 28978    &
                   &  0.9                     &  0.007   & 0.039               & 0.000              &  8123     \\    
    \hline
1902-1927          & -0.9                     & -0.304   & 0.014               & 0.013              & 64098    &
1977-2000          & -0.9                     & -0.103   & 0.085               & 0.005              &  1736   \\    
                   & -0.7                     & -0.486   & 0.016               & 0.021              & 48795    &
                   & -0.7                     & -0.018   & 0.104               & 0.001              &  1156    \\    
                   & -0.5                     & -0.490   & 0.017               & 0.021              & 43086    &
                   & -0.5                     & -0.472   & 0.120               & 0.021              &   842    \\    
                   & -0.3                     & -0.459   & 0.019               & 0.020              & 35549    &
                   & -0.3                     & -0.400   & 0.124               & 0.018              &   798     \\    
                   & -0.1                     & -0.383   & 0.020               & 0.017              & 29543    &
                   & -0.1                     & -0.514   & 0.115               & 0.023              &   919     \\    
                   &  0.1                     & -0.265   & 0.022               & 0.012              & 24759    & 
                   &  0.1                     & -0.439   & 0.117               & 0.019              &   892      \\    
                   &  0.3                     & -0.203   & 0.021               & 0.009              & 28268    &
                   &  0.3                     & -0.352   & 0.118               & 0.015              &   889     \\    
                   &  0.5                     & -0.160   & 0.019               & 0.007              & 33350    & 
                   &  0.5                     & -0.122   & 0.112               & 0.005              &   988     \\    
                   &  0.7                     & -0.102   & 0.018               & 0.004              & 37409    &
                   &  0.7                     & -0.196   & 0.112               & 0.000              &   994     \\    
                   &  0.9                     &  0.039   & 0.017               & 0.027              & 43954    &
                   &  0.9                     &  0.026   & 0.115               & 0.011              &  1125     \\    
    \hline
1927-1952          & -0.9                     & -0.345   & 0.016               & 0.015              & 50612    &
                   &                          &          &                     &                    &        \\    
                   & -0.7                     & -0.471   & 0.017               & 0.021              & 40536    &
                   &                          &          &                     &                    &         \\    
                   & -0.5                     & -0.500   & 0.019               & 0.022              & 33626    &
                   &                          &          &                     &                    &         \\    
                   & -0.3                     & -0.494   & 0.021               & 0.022              & 26747    &
                   &                          &          &                     &                    &          \\    
                   & -0.1                     & -0.411   & 0.025               & 0.018              & 19836    &
                   &                          &          &                     &                    &          \\    
                   &  0.1                     & -0.328   & 0.027               & 0.014              & 16775    & 
                   &                          &          &                     &                    &           \\    
                   &  0.3                     & -0.267   & 0.026               & 0.012              & 18598    &
                   &                          &          &                     &                    &          \\    
                   &  0.5                     & -0.177   & 0.026               & 0.008              & 18786    & 
                   &                          &          &                     &                    &          \\    
                   &  0.7                     & -0.113   & 0.025               & 0.005              & 19395    &
                   &                          &          &                     &                    &          \\    
                   &  0.9                     & -0.036   & 0.023               & 0.002              & 23043    &
                   &                          &          &                     &                    &          \\    
    \hline
    \end{tabular}
  \caption{\label{an0terrienresults} 
   Analysing powers $A_{N0}$ for the reaction ${\rm n}{\rm p} \rightarrow {\rm p}{\rm p} \pi^{-}$
   at $T_{\rm n}=550~{\rm MeV}$ for different bins in $M_{\rm pp}$.
   Quoted are statistical and systematic errors.}
  \end{center}
\end{table*}

\subsection{Results for small invariant proton-proton masses}
\noindent{
Since the zero-crossing in the forward direction at 550 MeV in Fig.\,\ref{terrien} 
is observed in the lowest $M_{\rm pp}$ regime only, the observed pattern is likely 
due to an interference between various partial waves with a pp($^{1}{\rm S}_{0}$) 
final state. To study this effect in more detail, analysing powers were determined 
by selecting events with small $M_{\rm pp}$ values. Since the reconstruction 
efficiency drops at small $M_{\rm pp}$, a loose cut, $M_{\rm pp} - 2\cdot M_{\rm p} < 6~{\rm MeV}$, 
was chosen in order to collect sufficient statistics. As a consequence, there 
is a significant dilution due to partial waves with the two protons being in a 
relative P-wave. Therefore, the results can be used only for a qualititative 
discussion.
}

\begin{figure}
\resizebox{0.5\textwidth}{!}{%
  \includegraphics{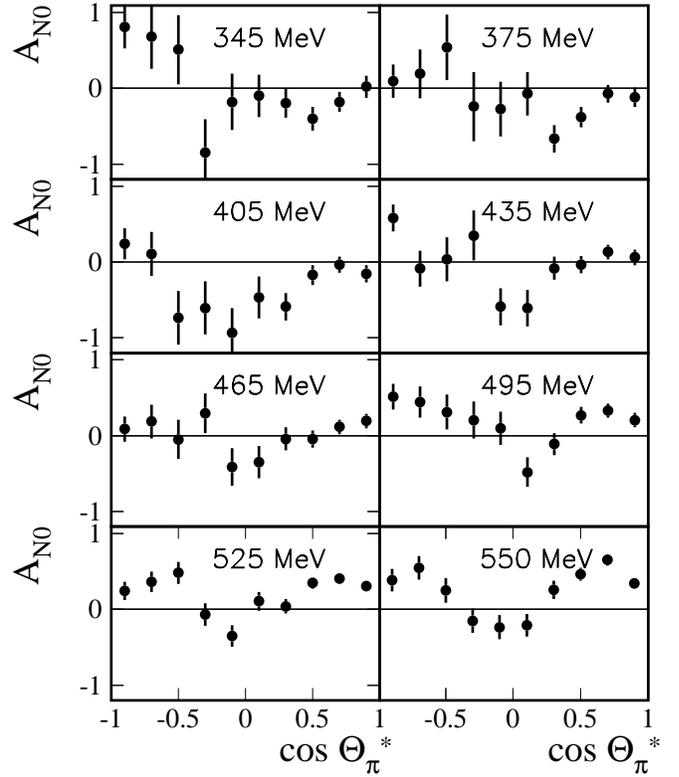}
}
\vspace{-0.0cm}       
\caption{
  Analysing powers $A_{N0}(\theta_{\pi}^{*})$ for small   
  proton-proton invariant masses 
  $M_{\rm pp}-2M_{\rm p} < 6~{\rm MeV}$. 
}
\label{mppsmall}      
\end{figure}

\noindent{
The $A_{N0}(\theta_{\pi}^{*})$-values for the small $M_{\rm pp}$ cut are shown 
in Fig.\,\ref{mppsmall}. Despite the loose $M_{\rm pp}$ cut there is only small 
statistics in the backward region since there the differential cross section and 
the reconstruction efficiency is smaller than in the forward region. Nevertheless, 
one can state that, in general, positive analysing powers are observed in the 
backward region and negative values around $\cos{\theta_{\pi}^{*}} \approx 0$. 
For beam energies above 405 MeV, positive analysing powers are observed in the 
forward direction the magnitude of which increases with neutron energy.
}

\noindent{
Two zero-crossings in $A_{N0}(\theta_{\pi}^{*})$ have already been reported at 
440 MeV\,\cite{DUNC2} and are interpreted as a contribution from Sd partial waves. 
A possible significant contribution from d-wave pions at quite small beam energies 
was also reported in a CELSIUS experiment measuring the reaction 
${\rm p}{\rm p} \rightarrow {\rm p}{\rm p} \pi^{0}$\,\cite{ZLO1}.
}

\noindent{
Given this, our data imply a relative increase of Sd partial 
waves as a function of beam energy. Such a behaviour fits the 
naive expectation for the energy dependence of partial waves. 
For a pp($^{1}{\rm S}_{0}$) final state, the excitation function 
is supposed to scale like $\eta^{2\cdot(\ell + 1)}$, see,
{\it e.g.}, Ref.\,\cite{DAU1}. Therefore, the contribution of 
d-wave pions should increase relative to s- or p-wave pions as 
the neutron energy increases. It should be noted however that 
this is only a qualitative argument. On one hand, the matrix 
element close to threshold shows a strong energy dependence 
in the pp-interaction part\,\cite{MEY1}. On the other hand, at 
very high energies, the momenta of the outgoing particles are 
large. As a consequence, the approximation entering this 
prediction, see, {\it e.g.}, Ref.\,\cite{HAN6}, is no longer 
justified. Moreover, a dynamical suppression of the Ss partial
wave is predicted in meson production models.
}

\begin{figure}
\resizebox{0.5\textwidth}{!}{%
  \includegraphics{fig.5}
}
\vspace{-0.0cm}       
\caption{
  Comparison of $A_{N0}(\cos{\theta_{\pi}^{*}})$ values ($\bullet$) 
  at 550 MeV subdivided in different bins of $M_{\rm pp}$ with 
  the results of Ref.\,\cite{TER1} at 572 MeV ($\circ$). 
}
\label{terrien}      
\end{figure}

\section{Conclusion}

\noindent{
The results of the analysing power $A_{N0}$ in the reaction 
$\vec{\rm n}{\rm p} \rightarrow {\rm p}{\rm p} \pi^{-}$ were 
preseneted as a function of the pion c.m. angle 
$\theta_{\pi}^{*}$ in different bins of the proton-proton 
invariant mass $M_{\rm pp}$ for neutron energies from threshold 
up to 570 MeV. Except for two experiments at 443 MeV\,\cite{BAC1} 
and 572 MeV\,\cite{TER1}, these are the first measurements of 
this observable over the full phase space and below the two-pion 
production threshold. The comparison with the reaction 
$\vec{\rm p}{\rm p} \rightarrow {\rm p}{\rm p} \pi^{0}$ 
clearly shows the presence of $\sigma_{01}$ in the reaction 
${\rm n} {\rm p} \rightarrow {\rm p} {\rm p} \pi^{-}$. 
The results obtained for small $M_{\rm pp}$ indicate a significant 
contribution from Sd partial waves at large neutron energies.
}

\noindent{
The additional knowledge from these spin dependent observables 
provides important information to disentangle the contributions 
from different partial waves. Such a partial wave analysis should 
be performed by combining data from both reactions,
${\rm n} {\rm p} \rightarrow {\rm p} {\rm p} \pi^{-}$ 
and $\vec{\rm p}{\rm p} \rightarrow {\rm p}{\rm p} \pi^{0}$.
Recently, such an analysis was performed for the $\sigma_{11}$
contribution using a complete set of polarisation observables 
measured in the reaction 
$\vec{\rm p}{\rm p} \rightarrow {\rm p}{\rm p} \pi^{0}$ for
proton beam energies between 315 MeV and 400 MeV\,\cite{MEY3}.
As a consequence, the $\sigma_{11}$ is already quite well known.
For the cross section $\sigma_{01}$, recent experimental results 
suggest that the main contribution in this energy region is 
provided by only two partial waves,
$^{3}{\rm D}_{1} \rightarrow ^{1}\!\!{\rm S}_{0}{\rm p}_{1}$ and
$^{3}{\rm S}_{1} \rightarrow \,^{1}{\rm S}_{0}{\rm p}_{1}$\,\cite{DAU1}
which will facilitate the analysis.
}

\noindent{
It would be also interesting to confront model calculations 
for pion production with the new data. However, for the 
reaction ${\rm n} {\rm p} \rightarrow {\rm p} {\rm p} \pi^{-}$,
except for very small proton-proton invariant masses, there 
are no published model calculations neither for differential 
cross sections nor for spin observables in the energy region
of interest.
}

\noindent{
For the elastic np scattering, $A_{00n0}$ was measured in 10 
energy bins over the backward hemisphere angular region. The 
results will improve the existing database for phase shift 
analyses.}

\section*{Acknowledgements}
\noindent{
We express our gratitude to C.~Lechanoine-Leluc, D.~Rapin 
and the DPNC of the University of Geneva for providing us 
the scintillators for the TOF wall and the associated 
electronics.
}

\noindent{
We thank M.~Laub and J.~Zicha for their help with the 
construction and the setup of the experiment.
}

\noindent{
We acknowledge the excellent cooperation with the 
staff of PSI during the installation and running 
of the experiment and the analysis as well. 
}

\noindent{
We appreciate the contribution of G.~Braun, R.~Fastner, 
H.~Fischer and J.~Urban.
}

\noindent{
We thank C.~Hanhart for the stimulating discussions.
}

\noindent{
This work has been funded by the German Bundesministerium 
f\"ur Bildung und Forschung under the contract No.~06FR845.
}

\end{document}